\documentclass[twocolumn,aps,prl,groupedaddress]{revtex4}
\usepackage{amssymb}
\usepackage{amsmath}
\usepackage{color}
\usepackage{graphicx}
\usepackage{marvosym}
\usepackage[normalem]{ulem}

\newcommand{\be}{\begin{equation}}
\newcommand{\ee}{\end{equation}}
\newcommand{\nn}{\nonumber\\}

\newcommand{\la}{\langle}
\newcommand{\ra}{\rangle}

\renewcommand{\vec}[1]{{\bf #1}}

\begin{document}
\title{Shift vector as the geometric origin of beam shifts}
\author{Li-kun Shi$^{1}$ and Justin C. W. Song$^{1,2}$}
\affiliation{$^1$Institute of High Performance Computing, Agency for Science, Technology, \& Research, Singapore 138632}
\affiliation{$^2$Division of Physics and Applied Physics, Nanyang Technological University, Singapore 637371}

\begin{abstract}
Goos-H\"{a}nchen (GH) and Imbert-Fedorov (IF) shifts are lateral and transverse displacements of a wavepacket reflecting off a surface. A dramatic real-space manifestation of wavepacket phases, they have traditionally been analyzed in a model dependent fashion. Here we argue that GH and IF shifts admit a general geometrical description and arise from a gauge invariant geometric phase. In particular, we show GH/IF shifts can be naturally captured by a shift vector, analogous to the shift vector from shift currents in the bulk photovoltaic effect. Employing Wilson loops to visualize the scattering processes contributing to the shift vector, we separate the shift into an intrinsic (depends solely on the system bulk) and an extrinsic part. This enables to establish a clear model-independent link between symmetry and the presence/absence of intrinsic and extrinsic GH/IF shifts.
\end{abstract}

\pacs{pacs}
\maketitle

Physical phenomena are typically insensitive to phases. This phase freedom embodies the ability to choose a coordinate system and is apparent in both classical and quantum systems. Departure from this rule of thumb, however, readily manifest in crystals with non-trivial Bloch overlaps. When integrated over a closed loop, wavepackets in such systems can acquire a gauge invariant Berry phase~\cite{Berry,Xiao} that manifests in a diversity of observables and intrinsic material properties including quantum oscillations~\cite{YuanboZhang}, topological edge/corner states~\cite{CLKane,Konig,Benalcazar,Schindler}, and the modern theory of polarization~\cite{Vanderbilt}.

\begin{figure}[t!] 
\includegraphics[width=1\columnwidth]{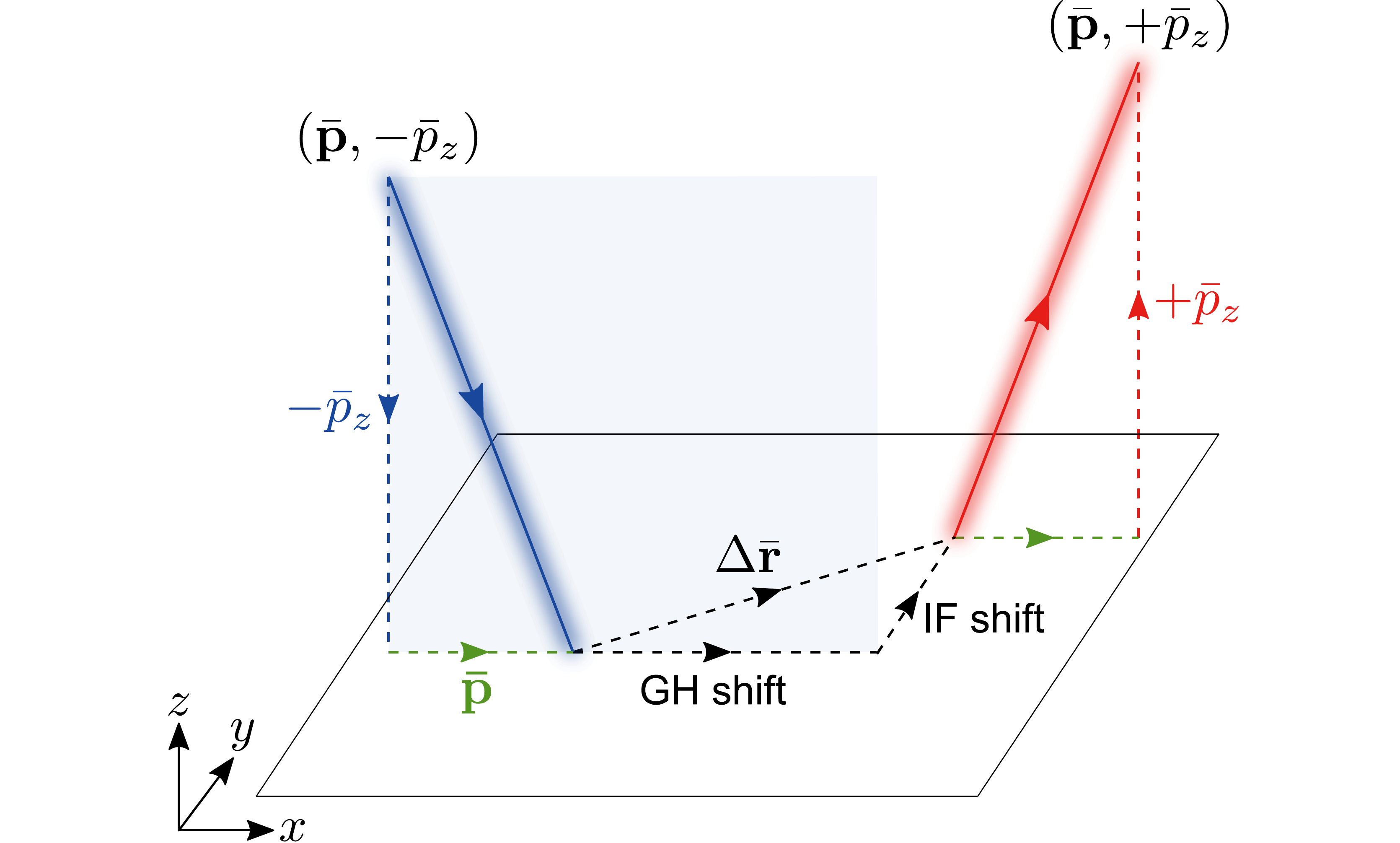}
\caption{Schematic Bloch beam shift $ \Delta \bar{\vec r} $ in real space upon total internal reflection. The Goos-H\"{a}nchen (GH) shift is parallel to the incident plane (blue shaded plane); the Imbert-Fedorov (IF) shift is perpendicular to the incident plane. Here the 3D incident wave vector is $( \bar{\vec p}, - {\bar p}_z )$. The reflecting boundary at $z = 0$ is parallel to $\bar{\vec p}$. As a result, $\bar{\vec p}$ remains conserved throughout the total internal reflection, while the wave vector perpendicular to the interface flips sign as $- {\bar p}_z \to + {\bar p}_z$.}
\label{fig1}
\end{figure}

Another striking example of phase sensitivity occurs when wavepackets reflect off a boundary. On reflection, wavepackets can acquire a reflection phase profile that leads to a real-space shift between the incident and reflected beam positions termed Goos-H\"anchen (GH) and Imbert-Federov (IF) shifts (Fig.~\ref{fig1}) defying the expectations of conventional ray optics~\cite{GoosHanchen,ImbertFedorov}. Such shifts can be found in many wave-media that range from free-space optics~\cite{Bliokh2013}, photonics~\cite{Onoda,Bliokh2015}, to electronics~\cite{Miller,XiChen,Beenakker,Wu,Yang,Jiang,Chattopadhyay,Liu1,Liu2,Shi}. GH/IF shifts have long been thought to be an ``extrinsic'' effect, and can be controlled by the particular superposition comprising the wavepacket beam~\cite{Bliokh2013,Onoda,Bliokh2015,Miller,XiChen,Beenakker,Wu} and boundary properties~\cite{Beenakker,Wu,Yang,Jiang,Chattopadhyay}. However, in topological media, these shifts have recently been found to be sensitive to the intrinsic characteristics of the bulk Bloch eigenstates. For example, in a topological Weyl semimetal, IF shifts flip in sign depending on which Weyl node is used to compose the incident wavepacket~\cite{Yang,Jiang}.

To what extent are GH/IF shifts intrinsic or extrinsic? Can GH/IF shifts also be described by a (unified) gauge invariant geometrical phase, and if so, which one? At first blush, such a geometrical description might seem counterintuitive given that GH/IF shifts seem to result from scattering between states over an {\it open} line path in state space (initial $\neq$ final). In contrast, gauge invariant Berry phases typically accumulate over {\it closed} loops (initial $=$ final). Furthermore, Berry phases are most commonly used to track the evolution of wavepackets in the intrinsic bulk, and do not typically involve a boundary.

Here we set out to address these questions, and argue that GH/IF shifts can be described in a completely geometrical fashion. In particular, we find that GH/IF shifts can generically be characterized by a gauge invariant shift vector that encodes both intrinsic (e.g., internal structure from Bloch band geometry of the periodic media) as well as extrinsic components (e.g., details of reflecting boundary) on the same footing. As we explain below, the twin roles of intrinsic and extrinsic components in GH/IF shifts naturally arise from a single gauge invariant geometric phase -- namely a Pancharatnam-Berry phase that tracks the scattering process -- effectively capturing both boundary and bulk Bloch eigenstate dependence. We note, parenthetically, that even though we focus on effects of Bloch band geometry (internal structure) in periodically structured media, our conclusions apply equally to beams with other forms of internal structure such as those from polarization/spinor degrees of freedom. 

Whilst providing an overall unified framework for understanding the origins of GH/IF shifts, this geometrical description can also be used as a powerful tool for analyzing the shifts in a model-independent way. As an illustration, we employ a pair of Wilson loops to separate the reflection process into individually distinct and gauge invariant intrinsic and extrinsic components. We find that the intrinsic contribution depends entirely on the Berry curvature and appear in systems with broken inversion and/or time-reversal symmetry. In contrast, while extrinsic components are generically non-zero, extrinsic IF shifts vanish in the presence of rotational symmetry. These provide clear symmetry conditions governing GH/IF shifts.

{\it Shift vector and GH/IF shifts ---} We begin by constructing wavepacket beams when they are incident (upon) and reflected (from) a boundary. We note that in experiments of (as well as proposals for observing) GH/IF shifts, wavepacket beams are typically constructed over a very narrow frequency range~\cite{Beenakker,Wu,Jiang,Yang,Wang,Shi,Liu1,Liu2,Chattopadhyay,Chen,Fei,He}. These produce real-space intensity patterns that are static and convenient to image. To replicate this, we consider wavepacket beams at a single frequency $\omega_0$. This constrains the superposition of Bloch eigenstates, that form the wavepacket beams, to lie on a surface of constant frequency.

Incorporating the above constraint, we write a Bloch beam incident from $ z = + \infty$ to $z = 0$ (see Fig.~\ref{fig1}) as 
\begin{equation}
\Psi^{\rm i} ( \vec r, z)  =
\int {\rm d} \vec p \, f (\vec p) u^{\rm i} (\vec p)  \exp [ i \vec p \cdot \vec r - i p_z (\vec p) z ]  ,
\label{eq:in}
\end{equation}
where $\vec p = (p_x,p_y)$ and $\vec r = (r_x,r_y)$ denote the wavevector and position in the $x$-$y$ plane, and we have written incident Bloch eigenstates with a fixed frequency $\omega_0$ as $u^{\rm i} (\vec p) \equiv u [ \vec p, - p_z ( \vec p ) ]$. Here $u (p_x, p_y, p_z)$ is the Bloch eigenstates without frequency constraint, which encodes the internal structure (or spinor texture) of the $z>0$ region either from the periodic lattice or from intrinsic spin-orbit interaction such as polarization~\cite{Onoda,Bliokh2015} or spin~\cite{Miller} degrees of freedom coupled to momentum, and $ - p_z (\vec p) < 0 $ denotes its incident direction as shown in Fig.~\ref{fig1}. We emphasize that, for a single frequency beam (or narrow frequency beam), $p_z (\vec p) > 0$ is a function of $ \vec p$, as is determined by the dispersion relation on the constant frequency surface $\omega_{(z>0)} [\vec p , p_z (\vec p) ]= \omega_0$. In Eq.~(\ref{eq:in}) and the following, we will consider a real incident distribution function $f (\vec p)$ that is well-peaked at $\bar{\vec p}$, i.e., the 3D incident wavevector is peaked at $(\bar{\vec p}, - \bar{p}_z)$ where $\bar{p}_z = p_z (\bar{\vec p})$ (see Fig.~\ref{fig1}). 

At the boundary $z = 0$, the incident beam undergoes a reflection. The reflected beam in $z > 0$ region can be written as 
\begin{equation}
\Psi^{\rm r}  (\vec r, z)  =  \int {\rm d} \vec p \, f (\vec p) r (\vec p) u^{\rm r} (\vec p) 
\exp [ i \vec p \cdot \vec r + i p_z (\vec p) z ]  ,
\label{eq:re}
\end{equation}
with reflected Bloch eigenstates $u^{\rm r} (\vec p) \equiv u [\vec p, + p_z (\vec p)] $, and reflection coefficient $r (\vec p)$ which relates the reflected and incident beam.

The reflection coefficient $r (\vec p)$ can be obtained by requiring continuity of the wavefunction at the boundary such that $ \Psi^{\rm i} (\vec r, 0^+) + \Psi^{\rm r} (\vec r, 0^+)  = \Psi^{\rm t} ( \vec r, 0^-) $~\cite{Beenakker,Wu,Yang,Jiang,Chattopadhyay}, where $\Psi^{\rm t} ( \vec r, z ) = \int {\rm d} \vec p f (\vec p) t (\vec p) w^{\rm t} (\vec p) \exp [ i \vec p \cdot \vec r - i p_z^{\rm t} (\vec p) z ] $ is the transmitted wave. $ w^{\rm t} (\vec p) \equiv w [ \vec p, p_z^{\rm t} (\vec p) ] $ is the eigenstate in the $z < 0$ region having the same frequency $\omega_0$. Here $t (\vec p)$ is the transmission coefficient. We note that similar to that discussed above, $p_z^{\rm t} (\vec p) > 0$ is also determined by the dispersion relation at the constant frequency $\omega_{(z<0)} [ \vec p , p_z^{\rm t} (\vec p) ] = \omega_0$. Wavefunction continuity leads to $u^{\rm i} (\vec p) + r (\vec p) u^{\rm r} (\vec p) = t (\vec p) w^{\rm t} (\vec p) $. By defining a unique auxiliary state vector $v (\vec p)$ perpendicular to $w^{\rm t} (\vec p)$, i.\,e., $\la v (\vec p) | w^{\rm t} (\vec p) \ra = 0$, we obtain a simple form for the reflection coefficient as $r (\vec p) = - \la v (\vec p) | u^{\rm i} (\vec p) \ra  \la v (\vec p) | u^{\rm r} (\vec p) \ra^{-1} $. 

For brevity and following previous work~\cite{Beenakker,Wu,Yang,Jiang,Chattopadhyay}, in the main text we focus on a  two-band system which possesses only a single reflected (transmitted) channel; $u^{\rm i,r} (\vec p)$ as well as $w^{\rm t} (\vec p)$ are two-component eigenvectors. The orthogonal requirement $\la v (\vec p) | w^{\rm t} (\vec p) \ra = 0$ uniquely determines the auxiliary state vector $ | v (\vec p) \ra $ [up to a $U(1)$ gauge that does not affect our conclusions, see Supplementary Information, {\bf SI}]. The uniqueness of $ | v (\vec p) \ra $, and the formal expression for the reflection coefficient $r (\vec p) = - \la v (\vec p) | u^{\rm i} (\vec p) \ra  \la v (\vec p) | u^{\rm r} (\vec p) \ra^{-1} $ are valid beyond two-band systems (see {\bf SI}).

When the incident frequency $\omega_0$ is within the gap of the medium in $z<0$ region, $p_z^{\rm t} (\vec p) \to - i \kappa_z^{\rm t} (\vec p) $ becomes imaginary making $w^{\rm t} (\vec p)$ an evanescent mode with a decay length $1/ \kappa_z^{\rm t} (\vec p)$. In this case, total internal reflection occurs, and the reflection coefficient has to be unitary. As a result, the reflection coefficient can be expressed as a pure phase $ r (\vec p) = \exp [ i \phi^{\rm r} (\vec p)]$ and reads as
\begin{equation}
\phi^{\rm r} (\vec p)  = \arg [ \la v (\vec p) | u^{\rm i} (\vec p) \ra \la u^{\rm r} (\vec p) | v (\vec p) \ra  ] + \pi ,
\label{eq:ReflectedPhase}
\end{equation}
where $\arg [ z ] $ denotes the polar angle (mod $2\pi$) of complex $z$, and we used the identity $\arg [z_1^{} z_2^{-1} ] = \arg [z_1^{} z_2^*]$. In what follows, we focus on total internal reflection. 

When the amplitude profile $f (\vec p)$ is sharply peaked around $\bar{\vec p}$ as found in wavepacket beams, the beam peak positions for both incident and reflected beam intensity profiles $|\Psi^{\rm i,r} (\vec r, z = 0)|^2$ on the $z = 0$ plane are obtained using standard stationary phase analysis~\cite{Supp}
\begin{equation}
\bar{\vec r}^{\rm i} = {\vec A}^{\rm i} (\bar{\vec p}) ,
\quad
\bar{\vec r}^{\rm r} = {\vec A}^{\rm r} (\bar{\vec p}) - \nabla_{\vec p} \phi^{\rm r} (\vec p ) |_{\bar{\vec p}},
\end{equation}
where ${\vec A}^{\rm i,r} (\bar{\vec p}) = \la u^{\rm i,r} (\vec p) | i \nabla_{\vec p} u^{\rm i,r} (\vec p) \ra_{\bar{\vec p}} $ is the Berry connection restricted to the constant frequency surface $\omega_{(z>0)} [ \vec p , p_z (\vec p) ]= \omega_0$. Evidently, the presence of the Berry connection indicates that the absolute positions of $\bar{\vec r}^{\rm i,r}$ are gauge variant. However, the {\it difference} between the positions $\bar{\vec r}^{\rm i}$ and $\bar{\vec r}^{\rm r}$ that encode the shift in position between incident and reflected beam (beam shift vector) are gauge invariant: 
\begin{equation}
\Delta \bar{\vec r} =   {\vec A}^{\rm r} (\bar{\vec p}) - {\vec A}^{\rm i} (\bar{\vec p}) - \nabla_{\vec p} \phi^{\rm r} (\vec p) |_{\bar{\vec p}}  ,
\label{eq:shift}
\end{equation}
This gauge invariance can be explicitly verified: for e.g., making an arbitrary gauge transformation $ u^{\rm i,r} (\vec p) \to u^{\rm i,r} (\vec p) \exp [i \chi^{\rm i,r} (\vec p)] $, we obtain ${\vec A}^{\rm i,r} (\vec p) \to {\vec A}^{\rm i,r} (\vec p) - \nabla_{\vec p} \chi^{\rm i,r} (\vec p)$. Similarly, using Eq.~(\ref{eq:ReflectedPhase}), we find that the reflection phase transforms as $\phi^{\rm r} (\vec p) \to \phi^{\rm r} (\vec p) - \chi^{\rm r} (\vec p) + \chi^{\rm i} (\vec p)$. As a result, changes to $ {\vec A}^{\rm i,r} (\vec p) $ and $ \nabla_{\vec p} \phi^{\rm r} (\vec p) $ under the gauge transformation cancel, leaving $\Delta \bar{\vec r}$ gauge invariant. 

The beam shift $ \Delta \bar{\vec r} $ includes both GH and IF shifts. The GH shift is parallel to the incident plane, and its magnitude is $ \delta {\bar r}_{\rm GH} = ( \bar{\vec p} \cdot \Delta \bar{\vec r} ) / |\bar{\vec p}| $; while the IF shift is perpendicular to the incident plane, whose magnitude is $ \delta {\bar r}_{\rm IF} = (\bar{\vec p} \times \Delta \bar{\vec r} ) \cdot \hat{z}/ |\bar{\vec p}| $ (see Fig.~\ref{fig1}).

On a physical level, the Berry connection $\vec A^{\rm i,r} (\bar{\vec p})$ in Eq.~(\ref{eq:shift}) is related to an intra-cell coordinate~\cite{Sundaram,Morimoto} for the incident and reflected Bloch beam intensities at $z = 0$ boundary. The difference between the two intra-cell coordinates $\vec A^{\rm r} (\bar{\vec p}) - \vec A^{\rm i} (\bar{\vec p})$, in addition to the gradient of phase difference $- \nabla_{\vec p} \phi^{\rm r} (\bar{\vec p})$, the conventional source of the beam shift, gives the full shift vector for beams possessing an internal structure. 

The shift vector in Eq.~(\ref{eq:shift}), appearing here in the context of representing GH and IF beam shifts (for the first time to our knowledge), echoes phenomena for Bloch waves in other contexts, for e.g., non-linear shift current in the bulk photovoltaic effect where a shift vector enters via induced inter-band transitions~\cite{Baltz,Young,Sipe,Morimoto,Chaudhary}, as well as side jumps found in anomalous Hall materials~\cite{Sinitsyn}. All of them can be expressed in terms of a shift vector [analogous to the beam shift vector that we discuss in Eq.~(\ref{eq:shift})] and have similar physical origins, namely the scattering between Bloch states that results in an intra-cell coordinate change, as well as a phase shift gradient which makes the overall shift vector gauge invariant.

\begin{figure}[t!] 
\includegraphics[width=1\columnwidth]{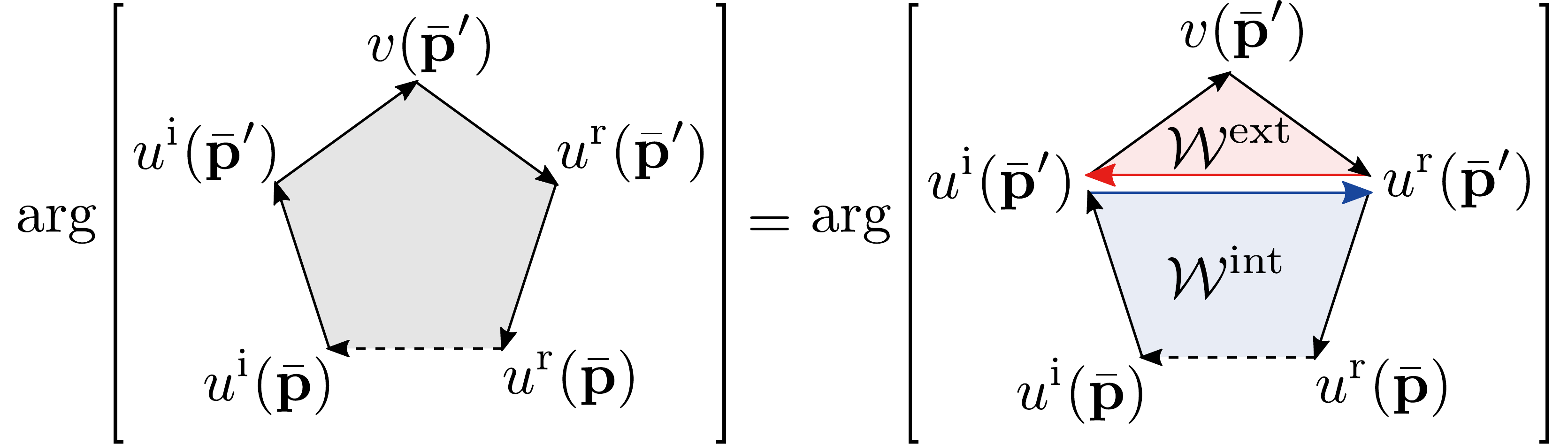}
\caption{Schematic showing the separation of intrinsic and extrinsic contributions into distinct (gauge invariant) Wilson loops ${\cal W}^{\rm int}$ (blue region loop) and ${\cal W}^{\rm ext}$ (red region loop). Connected solid black arrows denote the the scattering of state vectors from
$u^{\rm i} (\bar{\vec p}) \to u^{\rm i} (\bar{\vec p}') \to v (\bar{\vec p}') \to u^{\rm r} (\bar{\vec p}') \to u^{\rm r} (\bar{\vec p}) $, and dashed black arrows represent the extra $\vec q$-independent factor $\la u^{\rm r} (\bar{\vec p}) | u^{\rm i} (\bar{\vec p}) \ra$ which does not affect the shift vector. Together they form a closed Wilson loop ${\cal W} (\bar{\vec p}, \bar{\vec p}')$ (gray loop). Blue and red arrows on the right hand side cancel with each other since $\arg [z^{} \cdot z^* ] = 0 $.}
\label{fig2}
\end{figure}

{\it Wilson loops and intrinsic/extrinsic separation ---}
While expressed in terms of Berry connections and reflected phase gradients in Eq.~(\ref{eq:shift}), each part of the shift vector is still gauge dependent. Can the intrinsic and extrinsic contributions to the GH/IF shifts be separated in a gauge invariant way? Furthermore, since GH/IF shift are a phase sensitive phenomena, what geometric phase do they derive from? To address these, we note that both the reflection phase in Eq.~(\ref{eq:ReflectedPhase}) as well as the Berry connection can be captured by transitions between state vectors (Wilson lines/segments). Indeed, the Berry connection essentially encodes phases between different Bloch eigenstates $\la u^{\rm i,r} (\vec p) | u^{\rm i,r} (\vec p + \vec q) \ra = \exp [ - i \vec A^{\rm i,r} (\vec p) \cdot \vec q + {\cal O} (q^2) ] $, and can be expressed as $\vec A^{\rm i,r} (\vec p) = - \nabla_{\vec q} \arg [ \la u^{\rm i,r} (\vec p) | u^{\rm i,r} (\vec p + \vec q ) \ra ] |_{\vec q \to \vec 0}$. Using these, we rewrite the shift vector in Eq.~(\ref{eq:shift}) as
\begin{equation}
\Delta \bar{\vec r} = \nabla_{\vec q} \arg [ {\cal W} (  \bar{\vec p} , \bar{\vec p}') ] |_{\vec q \to \vec 0},
\quad
\bar{\vec p}' = \bar{\vec p} + \vec q,
\label{eq:shift-as-a-phase-gradient}
\end{equation}
where the (gauge invariant) Wilson loop 
\begin{align}
{\cal W} ( \bar{\vec p}, \bar{\vec p}' ) =\, & \la u^{\rm i} (\bar{\vec p}) | u^{\rm i} (\bar{\vec p}' ) \ra \la u^{\rm i} (\bar{\vec p}') | v (\bar{\vec p}')  \ra
\la v (\bar{\vec p}' ) | u^{\rm r} (\bar{\vec p}') \ra
\nn
& \cdot 
\la u^{\rm r} (\bar{\vec p}')  | u^{\rm r} (\bar{\vec p}) \ra \la u^{\rm r} (\bar{\vec p}) | u^{\rm i} (\bar{\vec p}) \ra,
\label{eq:Wilson-line}
\end{align}
encodes the scattering of state vectors from $u^{\rm i} (\bar{\vec p}) \to u^{\rm i} (\bar{\vec p}') \to v (\bar{\vec p}') \to u^{\rm r} (\bar{\vec p}') \to u^{\rm r} (\bar{\vec p}) $ (solid black lines in Fig.~\ref{fig2}). In obtaining Eq.~(\ref{eq:Wilson-line}) we have added $\la u^{\rm r} (\bar{\vec p}) | u^{\rm i} (\bar{\vec p}) \ra$ (last term) that is $\vec q$-independent; its contribution to $\Delta \bar{\vec r}$ vanishes under the action of $\nabla_{\vec q}$ in Eq.~(\ref{eq:shift-as-a-phase-gradient}). We note that even without the last term, the first four terms of Eq.~(\ref{eq:Wilson-line}) give a Wilson line that under the action of $\nabla_{\vec q}$ remains gauge invariant as $\bar{\vec p}'$ always appears in pairs.

$\arg [ {\cal W} (  \bar{\vec p} , \bar{\vec p}') ] $ is the (gauge invariant) Pancharatnam-Berry phase that describes the full scattering process for the GH/IF shift. We note that in free-space optics, gradients of a reflection phase are used to describe the GH shift~\cite{Bliokh2013,XiChen}. Eq.~(\ref{eq:shift-as-a-phase-gradient}) generalizes this notion to a {\it geometric} description: both GH and IF shifts are generically captured by gradients of a Pancharatnam-Berry phase. 

We now turn to separating out $\Delta \bar{\vec r}$ in Eq.~(\ref{eq:shift-as-a-phase-gradient}) into contributions that explicitly depend on the boundary [extrinsic, i.e. depends on $v (\vec p)$] and contributions that depend only on the bulk eigenstates [intrinsic, i.e. independent of $v (\vec p)$]. This can be done by noting that the argument of the Wilson loop $ {\cal W} (  \bar{\vec p} , \bar{\vec p}')$ in Eq.~(\ref{eq:Wilson-line}) can be decomposed into two smaller loops, $\arg [ {\cal W} (  \bar{\vec p} , \bar{\vec p}') ] = \arg [ {\cal W}^{\rm int} (\bar{\vec p}, \bar{\vec p}') \cdot {\cal W}^{\rm ext} (\bar{\vec p}') ]$ shown schematically as blue and red region loops in Fig.~\ref{fig2}.

The intrinsic Wilson loop (blue region loop, Fig.~\ref{fig2}) is
\begin{align}
{\cal W}^{\rm int} (\bar{\vec p}, \bar{\vec p}') = \, &  \la u^{\rm i} (\bar{\vec p}) | u^{\rm i} (\bar{\vec p}') \ra \la u^{\rm i} (\bar{\vec p}') | u^{\rm r} (\bar{\vec p}')  \ra
\nn
& \cdot \la u^{\rm r} (\bar{\vec p}')  | u^{\rm r} (\bar{\vec p}) \ra \la u^{\rm r} (\bar{\vec p}) | u^{\rm i} (\bar{\vec p}) \ra,
\end{align}
which is purely composed of Bloch eigenstates of the system bulk in the $z>0$ region. On the other hand, the extrinsic Wilson loop (red region loop in Fig.~\ref{fig2}) is
\begin{align}
{\cal W}^{\rm ext} (\bar{\vec p}') = \, & 
\la u^{\rm i} (\bar{\vec p}') | v (\bar{\vec p}') \ra
\la v (\bar{\vec p}') | u^{\rm r} (\bar{\vec p}') \ra
\nn
& \cdot \la u^{\rm r} (\bar{\vec p}')  | u^{\rm i} (\bar{\vec p}') \ra  ,
\end{align}
which contains information from the boundary, i.e., the auxiliary state vector $v (\vec p)$ which is perpendicular to the evanescent mode $w^{\rm t} (\vec p)$ at $z = 0$.

Using the identity $\arg[z_1 \cdot z_2] = \arg[z_1] + \arg[z_2]$, we obtain separate contributions from the intrinsic and extrinsic parts to the shift vector as
\begin{equation}
\Delta \bar{\vec r} = \nabla_{\vec q} \arg [ {\cal W}^{\rm int} (\bar{\vec p}, \bar{\vec p}') ] |_{\vec q \to \vec 0}
+ \nabla_{\vec p} \arg [ {\cal W}^{\rm ext} (\vec p ) ] |_{\bar{\vec p}},
\label{eq:SeparatedShift}
\end{equation}
where we have noted that ${\cal W}^{\rm ext} (\bar{\vec p}')$ depends solely on $\bar{\vec p}'$ so that the action of the gradient $\nabla_{\vec q} = \nabla_{{\vec p}'} = \nabla_{\vec p}$.

The above separation is physically meaningful, and as we show below, enable us to isolate contributions to the shift vector which are independent of the details of the boundary condition. To illustrate this and without losing generality, we consider an incident Bloch beam, whose incident wave vector $ ( \bar{\vec p} , - {\bar p}_z )  = ( {\bar p}, 0, - {\bar p}_z )$  is in the $z$-$x$ plane (see Fig.~\ref{fig1}): this gives a GH shift along the $x$-axis, and an IF shift along the $y$-axis.

\begin{figure}[t!] 
\includegraphics[width=1\columnwidth]{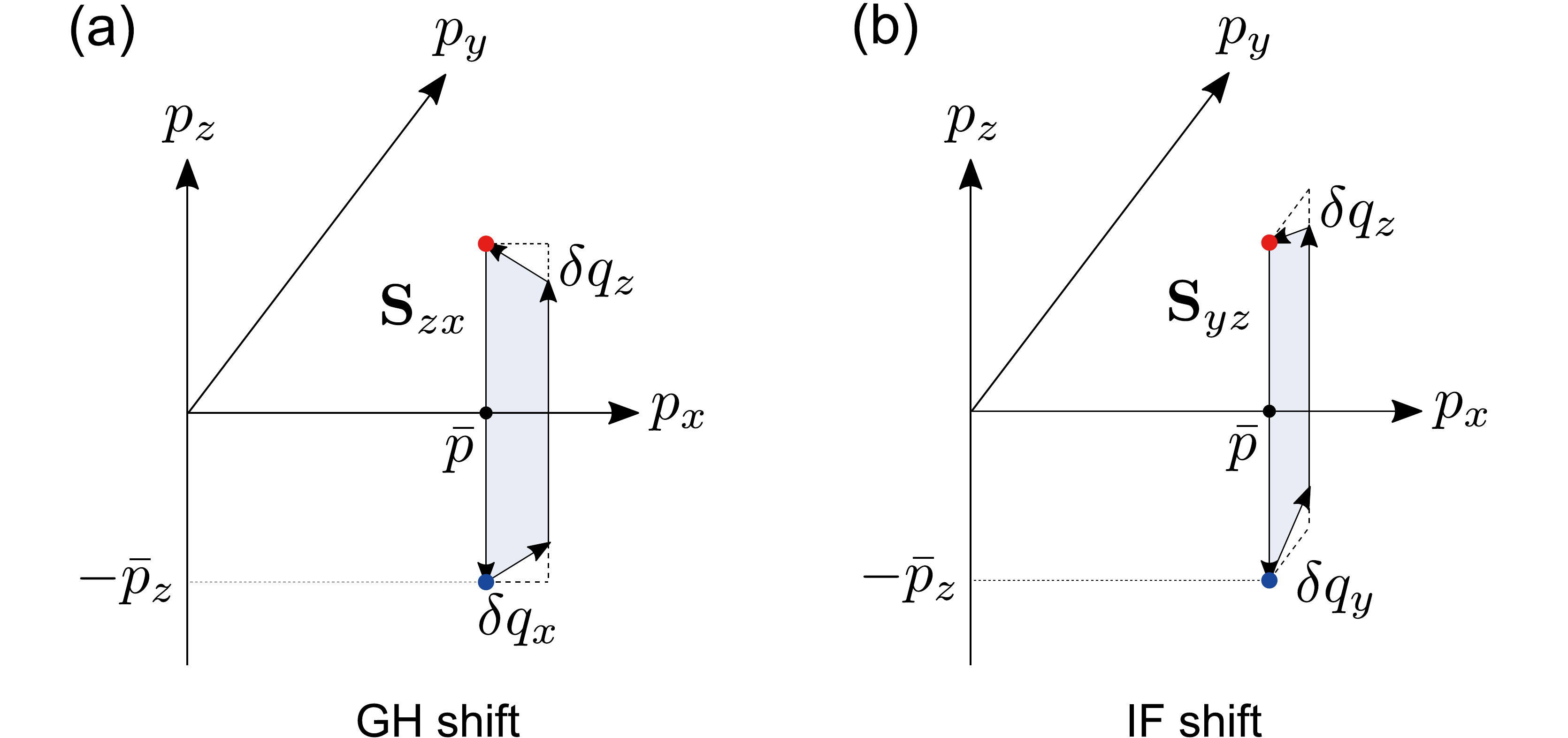}
\caption{The (blue) areas (a) $\vec S_{zx} $ and (b) $\vec S_{yz} $ enclosed by the intrinsic Wilson loop ${\cal W}^{\rm int} (\bar{\vec p}, \bar{\vec p} + \delta q_x ) $ and ${\cal W}^{\rm int} (\bar{\vec p}, \bar{\vec p} + \delta q_y ) $ in $\vec p$-space, which correspond to intrinsic GH and IF shifts illustrated in Fig.~\ref{fig1}, respectively. Blue (red) dot represents the peak incident (reflected) wavevector $(\bar{\vec p}, \mp \bar{p}_z)$. Wavevector for the intermediate state $u^{\rm i,r} (\bar{\vec p} + \delta q_{x(y)}) $  in 3D is $[ \bar{\vec p} + \delta q_{x (y)},  \mp p_z ( \bar{\vec p} + \delta q_{x (y)} ) ] = [ \bar{\vec p} + \delta q_{x (y)} , \mp ( \bar{p}_z - \delta q_z ) ] $ with $\delta q_z \propto \delta q_{x(y)}$.}
\label{fig3}
\end{figure}

We first focus on the intrinsic contribution. The GH (IF) shift from $ \nabla_{\vec q} \arg [ {\cal W}^{\rm int} (\bar{\vec p}, \bar{\vec p}') ] |_{\vec q \to \vec 0} $ in Eq.~(\ref{eq:SeparatedShift}) is
\begin{equation}
\Delta \bar{r}_{\rm GH (IF)}^{\rm int} =
{\rm lim}_{\delta q_{x (y)}  \to 0} 
\frac{ \arg [ {\cal W}^{\rm int} (\bar{\vec p}, \bar{\vec p} + \delta q_{x (y)} ) ] }
{ \delta q_{x (y)} }  ,
\label{eq:shift-as-a-dividing}
\end{equation}
with its Wilson loop parallel to the $z$-$x$ ($y$-$z$) plane enclosing an area $ \vec S_{zx}$ ($ \vec S_{yz}$) [see Fig.~\ref{fig3}a (Fig.~\ref{fig3}b)]. We note that the numerator in Eq.~(\ref{eq:shift-as-a-dividing}) is exactly the Pancharatnam-Berry phase $ \bar{\gamma}_{x (y)} \equiv \arg [ {\cal W}^{\rm int} ( \bar{\vec p}, \bar{\vec p} + \delta q_{x(y)} ) ] $ enclosed by the Wilson loop ${\cal W}^{\rm int} (\bar{\vec p} , \bar{\vec p} + \delta q_{x (y) } )$. Using Stoke's theorem, we can write the Berry phase as the Berry flux passing through $ \vec S_{zx}$ ($\vec S_{yz}$) in $\vec p$-space:
\begin{align}
\bar{\gamma}_{x (y)} & = 
\int_{ \vec S_{zx} (\vec S_{yz}) }
{\boldsymbol \Omega} (\vec p, p_z) \cdot {\rm d} \vec S
\nn
& = \mp \delta q_{x(y)} \int_{ - \bar{p}_z }^{ + \bar{p}_z }  {\rm d} p_z
\Omega_{y(x)} (\bar{\vec p}, p_z)  
+ {\cal O } ( \delta q_{x(y)}^2 ) ,
\label{eq:berry-flux}
\end{align}
where $ {\boldsymbol \Omega} (\vec p, p_z)$ is the Berry curvature for Bloch eigenstates $u (\vec p, p_z)$ in 3D, and the $-(+)$ sign comes from the orientation of $\vec S_{zx}$ ($\vec S_{yz}$). Applying Eq.~(\ref{eq:berry-flux}) into Eq.~(\ref{eq:shift-as-a-dividing}) yields the intrinsic GH and IF shifts as
\begin{equation}
\Delta \bar{r}_{\rm GH (IF)}^{\rm int} = \mp
\int_{ -\bar{p}_z }^{ + \bar{p}_z } {\rm d} p_z \,
\Omega_{y \, (x)} (\bar{\vec p}, p_z). 
\label{eq:BerryFluxIntrinsic}
\end{equation}
Interestingly, Eq.~(\ref{eq:BerryFluxIntrinsic}) dictates that for a system with zero Berry curvature (in the presence of both inversion and time-reversal symmetry), the intrinsic contribution to GH/IF shifts vanish, leaving only the extrinsic part.

While the extrinsic contribution to $\Delta \bar{\vec r}$ depends on details of the boundary, as we now discuss, there exist situations where its contribution to the IF shift vanishes. One such example occurs when the overall system (including the boundary) has continuous rotational symmetry in the $x$-$y$ plane, i.e., ${\cal W}^{\rm ext} ( \bar{\vec p} )  = {\cal W}^{\rm ext} ( \hat{R}_\theta \bar{\vec p} )$ in which $\hat{R}_\theta$ is a rotation matrix about the $z$-axis. For an infinitesimal rotation $\hat{R}_{ \delta \theta}$, the difference $(\hat{R}_{ {\rm \delta} \theta } \bar{\vec p}) - \bar{\vec p} = \delta q_y \hat{\vec{y}}$ is perpendicular to $\bar{\vec p}$. Applying this rotational symmetry to the extrinsic IF shift, we obtain 
\begin{equation}
\Delta \bar{r}_{\rm IF}^{\rm ext} =
{\rm lim}_{\delta \theta \to 0} 
\frac{ \arg [ {\cal W}^{\rm ext} ( \hat{R}_{ \delta \theta } \bar{\vec p} ) ] -  \arg [ {\cal W}^{\rm ext} ( \bar{\vec p} ) ] } { |\bar{\vec p}| \delta \theta }  = 0 ,
\end{equation}
that vanishes. Here the denominator $|\bar{\vec p}| \delta \theta$ is the magnitude of $(\hat{R}_{ \delta \theta } \bar{\vec p} ) - \bar{\vec p} $.

Therefore, in the presence of continuous rotational symmetry, we conclude that the total IF shift is solely determined by the intrinsic part in Eq.~(\ref{eq:BerryFluxIntrinsic}): $\Delta \bar{r}_{\rm IF} = \Delta \bar{r}_{\rm IF}^{\rm int}$. Indeed, a total $\Delta \bar{r}_{\rm IF}$ that follows Eq.~(\ref{eq:BerryFluxIntrinsic}) coincides with the shift expected from semiclassical equations of motion in Weyl semimetals~\cite{Jiang,Yang}. Our model-independent analysis, valid for general scattering including at a sharp interface (non-adiabatic process), unveils how the presence of rotational symmetry is key in ensuring $\Delta \bar{r}_{\rm IF}^{\rm ext}$ vanishes to yield a total $\Delta \bar{r}_{\rm IF}$ that is purely intrinsic~\footnote{Our conclusion that the absence of $\delta \bar{r}_{\rm IF}^{\rm ext}$ is guaranteed by rotational symmetry for any $\bar{\vec p}$, does not contradict with accidental vanishing of $\Delta \bar{r}_{\rm IF}^{\rm ext} $ when ${\cal W}^{\rm ext} (\bar{\vec p}) $ reaches certain extremal points in $\vec p$-space.}. When rotational symmetry is broken, total $\Delta \bar{r}_{\rm IF}$ generically departs from Eq.~(\ref{eq:BerryFluxIntrinsic}).

The extrinsic GH shift $\Delta \bar{r}_{\rm GH} $ is in general non-zero due to the lack of scale invariance $ {\cal W}^{\rm ext} ( \hat{L}_\lambda \bar{\vec p} ) \neq {\cal W}^{\rm ext} ( \bar{\vec p} ) $, where $\hat{L}_\lambda \bar{\vec p} \equiv (1+\lambda)\bar{\vec p} $. Interestingly, for a single ideal Weyl node, which is a monopole of its Berry curvature, ${\boldsymbol \Omega} (\vec p, p_z)$ is always parallel to $(\vec p, p_z)$. In this case, $\Omega_y (\bar{p}, 0, p_z) = 0$ and $\Delta \bar{r}_{\rm GH}^{\rm int} = 0$, i.e., the GH shift is solely contributed by the extrinsic part $ \Delta \bar{r}_{\rm GH} = \Delta \bar{r}_{\rm GH}^{\rm ext} $.

Our work demonstrates how GH and IF shifts can be generally described in a purely geometric fashion and arises from a gauge invariant Pancharatnam-Berry phase; it naturally captures contributions from both the reflecting boundary as well as the bulk (internal structure-related) in which the wavepacket propagates in. This generalized description can also be used as a powerful tool for analyzing GH/IF shifts, enabling us to separate and identify (for the first time to our knowledge) the role of intrinsic and extrinsic contributions and establish clear symmetry requirements for their existence in a model-free way. Perhaps most exciting is the deep connection between GH/IF shifts and a range of other phenomena that rely on the shift vector~\cite{Morimoto,Baltz,Sipe,Young,Chaudhary,Sinitsyn}. Given this shared geometrical connection, we anticipate the tools developed to analyze GH/IF shifts here can be readily employed to study a host of other types of shift vector phenomena such as the shift current in the bulk photovoltaic effect.

{\it Acknowledgements ---}
We gratefully acknowledge useful conversations with Yidong Chong. This work was supported by the Singapore National Research Foundation (NRF) under NRF fellowship award NRF-NRFF2016-05, a Nanyang Technological University start-up grant (NTU-SUG), and Singapore MOE Academic Research Fund Tier 3 Grant MOE2018-T3-1-002.

\clearpage
\newpage

\renewcommand{\theequation}{S-\arabic{equation}}
\renewcommand{\thefigure}{S-\arabic{figure}}
\renewcommand{\thetable}{S-\Roman{table}}
\makeatletter
\renewcommand\@biblabel[1]{S#1.}
\setcounter{equation}{0}
\setcounter{figure}{0}

\twocolumngrid

\section*{Supplementary Information for\\
``Shift vector as the geometric origin of beam shifts''}

\subsection{Wavepacket peaks from stationary phase analysis} 

In this section, we detail the standard method of stationary phases used to determine the wavepacket center. Using Eqs.~(\ref{eq:in}), (\ref{eq:re}), and (\ref{eq:ReflectedPhase}) in the main text, and expressing the Bloch overlaps as $\la u^{\rm i,r} (\vec p) | u^{\rm i,r} (\vec p + \vec q) \ra = \exp [ - i \vec A^{\rm i,r} (\vec p) \cdot \vec q + {\cal O} (q^2) ] $,
the intensity profiles $|\Psi^{\rm i,r} (\vec r, z = 0)|^2$ on the $z = 0$ plane
can be expressed as
\begin{equation}
|\Psi^{\rm i,r} (\vec r, 0)|^2 = \int  {\rm d} \vec p {\rm d} \vec q \, W (\vec p , \vec q)  \exp [ i \theta^{\rm i,r} ( \vec p, \vec q, \vec r) ] ,
\label{eq:intensity}
\end{equation}
where $W (\vec p, \vec q) = f (\vec p) f (\vec p + \vec q) 
\exp [ {\cal O} ( q^2) ] $ is a composite amplitude profile, and  
\begin{equation}
\theta^{\rm i} (\vec p , \vec q, \vec r)  
= \vec q \cdot [ \vec r - \vec A^{\rm i} (\vec p) ] + {\cal O} (q^2) ,
\end{equation}
\begin{equation}
\theta^{\rm r} (\vec p , \vec q, \vec r)  
= \vec q \cdot [ \vec r - \vec A^{\rm r} (\vec p) ] 
+ \phi^{\rm r} (\vec p + \vec q) - \phi^{\rm r} (\vec p) + {\cal O} (q^2) ,
\label{eq:theta}
\end{equation}
are composite phase factors that include both phase information of the superposition of plane waves comprising the beam, as well as the Berry connection
$\vec A^{\rm i,r} (\vec p) = \la u^{\rm i,r} (\vec p) | i \nabla_{\vec p} u^{\rm i,r} (\vec p) \ra $.

When the amplitude profile $f (\vec p)$ is sharply peaked around $\bar{\vec p}$, the composite amplitude profile $W (\vec p,\vec q)$ is similarly sharply peaked around $(\bar{\vec p}, \vec 0)$. This can be verified by noting $ \nabla_{\vec p} W (\vec p,\vec q) |_{( {\bf \bar p} , \vec 0)} = \nabla_{\vec q} W (\vec p,\vec q) |_{( {\bf \bar p} , \vec 0)} = 0 $. 

The (real-space) peak position $\bar{\vec r}^{\rm i,r}$ of $|\Psi^{\rm i,r} (\vec r, z = 0)|^2$ can then be directly determined by the standard stationary phase analysis applied onto Eq.~(\ref{eq:intensity}). In particular, this requires that
\begin{equation}
\nabla_{\vec q}  \theta^{\rm i, r} [ \bar{\vec p}, \vec q, \bar{\vec r}^{\rm i, r} ] |_{\vec q \to \vec 0 } = 0
\end{equation}
yielding a (real-space) peak position for $|\Psi^{\rm i,r} (\vec r, z = 0)|^2$ as
\begin{equation}
\bar{\vec r}^{\rm i} = {\vec A}^{\rm i} (\bar{\vec p}) ,
\quad
\bar{\vec r}^{\rm r} = {\vec A}^{\rm r} (\bar{\vec p}) - \nabla_{\vec p} \phi^{\rm r} (\vec p ) |_{\bar{\vec p}} .
\end{equation}
This demonstrates that the peak position of Bloch wavepackets depend on both the internal structure of Bloch eigenstates [encoded in ${\vec A}^{\rm i,r} (\bar{\vec p})$] as well as the superposition phases $\phi (\vec p)$. We note, parenthetically, that $\bar{\vec r}$ is gauge variant and captures the intra-cell coordinate~\cite{Sundaram,Morimoto} for a Bloch wave packet, arising from linear combinations of Bloch eigenstates in the unit cell. However, as discussed in the main ext, the difference between $\bar{\vec r}^{\rm i}$ and $\bar{\vec r}^{\rm r}$ is gauge invariant.

\subsection{Uniqueness of auxiliary state vector $ | v (\vec p) \ra $} 

In the main text, we showed that the wavefunction continuity requirement at the boundary leads to $ u^{\rm i} (\vec p) + r (\vec p) u^{\rm r} (\vec p) = t (\vec p) w^{\rm t} (\vec p) $, where $u^{\rm i,r} (\vec p)$ and $w^{\rm t} (\vec p)$ are two-component eigenvectors. This equation corresponds to two-band systems~\cite{Beenakker,Wu,Yang,Jiang,Chattopadhyay} which has a single reflected (transmitted) channel for a fixed incident frequency. In this case, assuming the transmitted mode is $ | w^{\rm t} (\vec p) \ra = [ w^{\rm t, (1)} (\vec p), w^{\rm t, (2)} (\vec p) ]^T $, then the unique auxiliary state vector is $ \la v (\vec p) | = [ w^{\rm t, (2)} (\vec p), - w^{\rm t, (1)} (\vec p) ]$, as can be readily verified: $\la v (\vec p) | w^{\rm t} (\vec p) \ra = 0$. Using the auxiliary state vector $v (\vec p)$, the reflection coefficient reads $r (\vec p) = - \la v (\vec p) | u^{\rm i} (\vec p) \ra  \la v (\vec p) | u^{\rm r} (\vec p) \ra^{-1} $.

We note that one can certainly choose an arbitrary $U (1)$ gauge for $| v (\vec p) \ra$, but this will not affect the physical result of the shift vector $\Delta \bar{\vec r}$, because the auxiliary state vector appears in pairs as $| v (\vec p) \ra \la v(\vec p) | $ in the Wilson loop [see Eq.~(\ref{eq:Wilson-line}) in the main text].

Below we use a specific example~\cite{Chattopadhyay} to concretely illustrate this procedure. We emphasize, however, that the formula for $r(\vec p)$ and the shift in the main text are general. We proceed by considering the following two-band model:
\begin{equation}
H =
\begin{bmatrix}
p_y & (p_x^2 - m_z) - i p_z \\
(p_x^2 - m_z) + i p_z & - p_y
\end{bmatrix},
\end{equation}
in which
\begin{equation}
m_z=
\begin{cases}
+ m_0 > 0 , & z > 0,~\text{Weyl media,}  \\
- m_1 < 0 ,  & z < 0,~\text{gapped media.}
\end{cases}
\end{equation}

Inside the Weyl media where $z>0$, the dispersion relations for two bands are $\pm [(p_x^2 - m_0)^2 + p_y^2 + p_z^2]$. If we focus on the positive branch, and fix the incident energy to be $\epsilon_0 > m_0 > 0$, the eigenvector $| u^{\rm i, r} (\vec p) \ra$ reads
\begin{equation}
| u^{\rm i,r} (\vec p) \ra =
\begin{pmatrix}
\cos [ \theta^{\rm i, r} (\vec p) / 2 ] \\
\sin  [ \theta^{\rm i, r} (\vec p) / 2 ] \exp[ i \phi^{\rm i, r} (\vec p) ] 
\end{pmatrix} ,
\end{equation}
where $ \cos [ \theta^{\rm i, r} (\vec p) ] = p_y / \epsilon_0 $, $\tan [ \phi^{\rm i, r} (\vec p) ] = \mp p_z (\vec p) /  (p_x^2 - m_0) $, and $p_z (\vec p) = [ \epsilon_0^2 - (p_x^2 - m_0)^2 - p_y^2 ]^{1/2}$. We note that the negative branch with a dispersion $ - [(p_x^2 - m_0)^2 + p_y^2 + p_z^2]$ cannot support a reflected mode at the incident energy $\epsilon_0 > 0$.

In the gapped media where $z  < 0$, the dispersion relation for the two bands are $\pm [(p_x^2 + m_0)^2 + p_y^2 + p_z^2]$. The transmitted eigenvector $| w^{\rm t} (\vec p) \ra$ with energy $\epsilon_0 > 0$ reads
\begin{equation}
| w^{\rm t} (\vec p) \ra =
\begin{pmatrix}
\cos [ \theta^{\rm t} (\vec p) / 2 ] \\
\sin  [ \theta^{\rm t} (\vec p) / 2 ] \exp[ i \phi^{\rm t} (\vec p) ]  
\end{pmatrix} ,
\end{equation}
where $ \cos [ \theta^{\rm t} (\vec p) ] = p_y / \epsilon_0 $, $\tan [ \phi^{\rm t} (\vec p) ] = - p_z^{\rm t} (\vec p) /  (p_x^2 + m_1) $, and $p_z^{\rm t} (\vec p) = [ \epsilon_0^2 - (p_x^2 + m_1)^2 - p_y^2 ]^{1/2}$. Again, we note that the negative branch with a dispersion $ - [(p_x^2 + m_0)^2 + p_y^2 + p_z^2]$ can not support any transmitted mode at the incident energy $\epsilon_0 > 0$.

In this case, the auxiliary state vector reads
\begin{equation}
\la v (\vec p) | =
\big( \sin  [ \theta^{\rm t} (\vec p) / 2 ] \exp[ i \phi^{\rm t} (\vec p) ], 
- \cos [ \theta^{\rm t} (\vec p) / 2 ] \big) ,
\end{equation}
which can be used directly in the formula for the reflection coefficient: $r (\vec p) = - \la v (\vec p) | u^{\rm i} (\vec p) \ra  \la v (\vec p) | u^{\rm r} (\vec p) \ra^{-1} $. We note that when $\epsilon_0 < m_1$, $p_z^{\rm t} (\vec p) \to - i \kappa^{\rm t} (\vec p)$ becomes imaginary and total reflection occurs.

This procedure also applies to models with a higher number of bands. For example, four-band models: in Refs.~\cite{Liu1,Liu2}, the authors studied the Andreev reflections at metal/superconductor interfaces, the wavefunction continuity requirement at the interface leads to $ \psi^{e+}+r \psi^{e-}+r_{A} \psi^{h-} = t_+ \psi_+^{\rm S}+t_- \psi_-^{\rm S} $, where $\psi$'s are four-component state vectors, while $r$'s and $t$'s are reflection and transmission coefficients for two reflected and two transmitted channels. In this case, to extract $r_A$, a unique four-component auxiliary state vector $|v \ra$ can be constructed using a Gram-Schmidt process, by satisfying the orthogonal relation $\la v | \psi^{e-} \ra = \la v | \psi_+^{\rm S} \ra = \la v | \psi_-^{\rm S} \ra = 0$.

For both the two-band \cite{Beenakker,Wu,Yang,Jiang,Chattopadhyay} and four-band \cite{Liu1,Liu2} examples, their wavefunction continuity requirements follow a general form:
\begin{equation}
u^{\rm i} (\vec p) + \sum_{\mu = 1}^{n_1} r_\mu (\vec p) u_\mu^{\rm r} (\vec p) = \sum_{\nu = 1}^{n_2} t_\nu (\vec p) w_\nu^{\rm t} (\vec p)  ,
\label{eq:GeneralizedBC}
\end{equation}
where $u$'s and $w$'s are $N$-component state vectors, $r$'s and $t$'s are reflection and transmission coefficients, and $n_1 + n_2 = N$ counts the total number of reflected and transmitted channels allowed by energy conservation, with $N= 2 $ or $N=4$ for two-band or four-band models. This reflects the fact that $N$ linear equations [Eq.~(\ref{eq:GeneralizedBC})] can determine $N$ unknown variables $\{ r_\mu (\vec p), t_\nu (\vec p) \}$, where $\mu (\nu) = 1, \dots, n_{1(2)}$. To extract $r_i (\vec p)$ that we are interested in, we can construct the unique state vector $| v (\vec p) \ra$ using the Gram-Schmidt process described above by satisfying the orthogonal relation $\la v (\vec p) | u_{\mu \neq i }^{\rm r} (\vec p) \ra = \la v (\vec p) | w_\nu^{\rm t} (\vec p) \ra = 0$ where $\mu (\nu) = 1, \dots, n_{1(2)}$.

We expect this scheme for constructing a unique auxiliary state vector $| v (\vec p) \ra$ to obtain the reflection (or indeed, the transmission) coefficient can be readily extend to more general cases, especially for models with an equal number $N/2$ for both conduction bands and valence bands. In this case, the total number of reflected and transmitted channels allowed by energy conservation is $N$, which is the same as the number of linear equations provided by the wavefunction continuity requirement.

\clearpage
\end{document}